\documentclass[a4paper,11pt]{article}
\pdfoutput=1 

\usepackage{jheppub} 

\usepackage[T1]{fontenc} 
\usepackage{psfrag}

\title{\boldmath Colorful plane vortices and Chiral Symmetry Breaking in $SU(2)$ Lattice Gauge Theory}

\author[a]{Seyed Mohsen Hosseini Nejad,}
\author[b]{Manfried Faber}
\author[b,c]{and Roman H\"ollwieser}

\affiliation[a]{Department of Physics, University of Tehran, P.O. Box 14395-547, Tehran 1439955961, Iran}
\affiliation[b]{Institute of Atomic and Subatomic Physics, Vienna University of Technology, Wiedner Hauptstr. 8-10, 1040 Vienna, Austria}
\affiliation[c]{Department of Physics, New Mexico State University, P.O. Box 30001, Las Cruces, NM 88003-8001, USA}

\emailAdd{smhosseininejad@ut.ac.ir}
\emailAdd{faber@kph.tuwien.ac.at}
\emailAdd{hroman@kph.tuwien.ac.at}

\abstract{We investigate plane vortices with color structure. The topological
charge and gauge action of such colorful plane vortices are studied in the
continuum and on the lattice. These configurations are vacuum to vacuum
transitions changing the winding number between the two vacua, leading to a topological charge $Q=-1$ in the continuum. After growing temporal extent of these vortices, the lattice topological charge approaches $-1$ and the index theorem is fulfilled. We analyze the low lying modes of the overlap Dirac operator in the background of these colorful plane vortices and compare them with those of spherical vortices. They show characteristic properties for spontaneous chiral symmetry breaking.}


\def\gtwid{{\,\raise.3ex\hbox{$>$\kern-.75em\lower1ex\hbox{$\sim$}}\,}}

\begin{document}
\maketitle
\flushbottom
\section{Introduction}

We know since many years~\cite{Savvidy:1977as} that the QCD-vacuum is non-trivial and has magnetic properties. Center vortices \cite{'tHooft:1977hy,Vinciarelli:1978kp,Yoneya:1978dt,Cornwall:1979hz,Mack:1978rq,Nielsen:1979xu} which are quantized magnetic flux tubes are very successful in explaining these magnetic properties leading to confinement, as indicated by numerical simulations \cite{DelDebbio:1996mh,Kovacs:1998xm,Engelhardt:1999wr,Bertle:2002mm,Engelhardt:2003wm,Hollwieser:2014lxa,Altarawneh:2014aa}. Recent results in SU(2) gauge theory~\cite{Greensite:2014gra} have also suggested that the center vortex model of confinement is more consistent with lattice results than other currently available models. Also in SU(3) gauge theory~\cite{Trewartha:2015ida} the long-range structure is contained within the center vortex degrees of freedom. 
In addition, numerical simulations have shown that center vortices could also account for phenomena related to chiral symmetry, such as topological charge~\cite{Bertle:2001xd,Engelhardt:2000wc,Engelhardt:2010ft,Hollwieser:2010mj,
Hollwieser:2011uj,Schweigler:2012ae,Hollwieser:2012kb,Hollwieser:2014mxa,Hollwieser:2015koa,Altarawneh:2015bya,Hollwieser:2015qea}
and spontaneous chiral symmetry breaking
\cite{deForcrand:1999ms,Alexandrou:1999vx,Engelhardt:1999xw,Reinhardt:2000ck2,Engelhardt:2002qs,Leinweber:2006zq,Bornyakov:2007fz,Hollwieser:2008tq,Hollwieser:2009wka,Bowman:2010zr,Hollwieser:2013xja,Brambilla:2014jmp,Hollwieser:2014osa,Trewartha:2014ona,Trewartha:2015nna}, as explained in the following.

The importance of center vortices is based in the competition between action and entropy in the euclidean path integral. The Boltzmann factor $\exp\{-S\}$ favors trivial vacua characterized by their winding number, whereas the measure favors configurations with large entropy. In gauge theories with center symmetry the entropy of field configurations is increased by non-trivial center transformations. Such transformations are applied to the set of temporal links in a given time-slice (or $x_i$ links in an $x_i$-slice). Restricting these center transformations to a three-dimensional ``Dirac'' volume produces a vortex, the surface of the Dirac volume. If the Dirac volume extends over the whole range of two coordinates with periodic boundary conditions this surface consists of two disconnected pieces. Center vortices get their importance form the fact that they contribute to the action at the surface of the Dirac volume only. By smoothing the transition between center elements from inside to outside of the Dirac volume the action contribution of vortices is decreasing. Due to the large entropy of restricted center transformations the QCD vacuum is crowded with center vortices of random structure, as Monte-Carlo calculations show~\cite{Bertle:1999tw}. Vortices as two-dimensional surfaces can pierce areas surrounded by Wilson loops and lead to their area law behavior. For low piercing probability the QCD string tension gets proportional to this probability. The presence and structure of center vortices guarantees the center symmetry of the confined phase. In the deconfined phase center vortices get aligned in time direction~\cite{Engelhardt:1999wr} explaining the loss of center symmetry. The shape of vortex surfaces influences the contribution of vortices to the topological charge~\cite{Reinhardt:2000ck,Engelhardt:2000wc,Bertle:2001xd,Engelhardt:2010ft,Hollwieser:2010mj,Hollwieser:2011uj} via intersections and writhing points. Changes of the orientations of vortex surfaces can be interpreted as monopole lines and have their origin in the color structure of vortices as shown in~\cite{Schweigler:2012ae,Hollwieser:2012kb,Hollwieser:2014mxa,Hollwieser:2015koa} for colorful spherical vortices. In~\cite{Schweigler:2012ae,Hollwieser:2012kb} we argued that all objects carrying topological charge contribute to the density of near zero-modes of the fermionic determinant. According to the Banks-Casher relation~\cite{Banks:1979yr} these modes are responsible for a finite chiral condensate, the order parameter of spontaneous chiral symmetry breaking. All these properties indicate that the most important non-perturbative properties of the QCD-vacuum can be understood within the center vortex model.

In this article we show that a non-trivial color structure carrying topological charge can also be implemented in plane vortices. 
In section \ref{sec:original} pairs of plane vortices, where one of the vortices is colorful, are constructed on the lattice. Then smoothed continuum field configuration of plane vortices are considered. Their action and topological charge are studied in sections \ref{sec:action} and \ref{sec:topdencont}. In section \ref{sec:gen_latt}, smoothed plane vortices on the lattice are derived from the continuum form. 
In the background of the colorful plane vortices we compute the eigenmodes of
the overlap Dirac operator in section \ref{sec:Dirac} and compare them with those for spherical vortices with the same topological charge and trivial gauge fields. In the last step in section~\ref{sec:conclusion} we summarize the main points of our study.
The results clearly show that colorful plane vortices produce the characteristic properties for chiral symmetry breaking.

\section{Plane vortex pairs with one colorful vortex}\label{sec:original}
In this section plane vortices with color structure are introduced. The
construction of unicolor plane vortices was introduced in \cite{Jordan:2007ff}
and explained in detail in \cite{Hollwieser:2011uj}. Their links vary in a
$U(1)$ subgroup of $SU(2)$, characterized by one of the Pauli matrices
$\sigma_i$, {\it i.e.}, $U_\mu=\exp(\mathrm{i} \alpha \sigma_i)$. E.g. for
$xy$-vortices $\mu=t$ links are nontrivial in one $t$-slice only, where they
vary with $z$. Since we use periodic boundary conditions for gauge fields,
vortices come in pairs of parallel sheets. The vortex sheets have thickness $2d$
around $z_1$ and $z_2$, in these regions the links differ from center elements. The orientation of the vortex flux is determined by the gradient of the angle $\alpha$, which we choose as a linear function of $z$, a coordinate perpendicular to the vortex
\begin{equation}\label{eq:phi-pl0}
\alpha_1(z)=\begin{cases}2\pi\\\pi\left[ 2-\frac{z-(z_1-d)}{2d}\right]\\
             \pi\\\pi\left[1-\frac{z-(z_2-d)}{2d}\right]\\0\end{cases}
\hspace{-4mm}\ldots,\quad
\alpha_2(z)=\begin{cases}0&0<z\leq z_1-d,\\
                \frac{\pi}{2d}[z-(z_1-d)]&z_1-d< z\leq z_1+d,\\
                \pi&z_1+d<z\leq z_2-d,\\
                \pi\left[1-\frac{z-(z_2-d)}{2d}\right]&z_2-d<z\leq z_2+d,\\
                0&z_2+d<z\leq N_z.\end{cases}
\end{equation}
\begin{figure}[h]
\centering
a)\includegraphics[width=0.46\columnwidth]{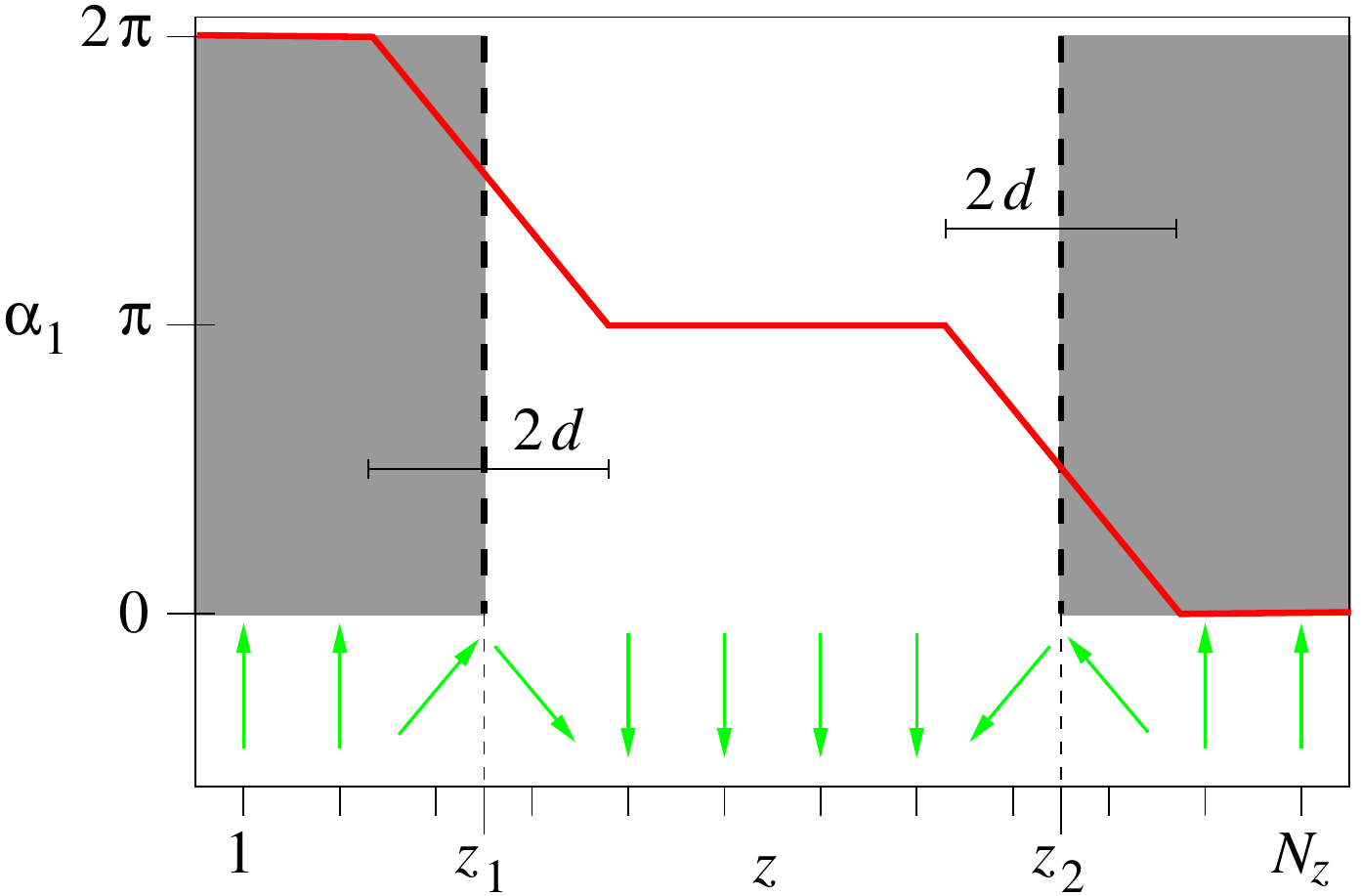}
b)\includegraphics[width=0.46\columnwidth]{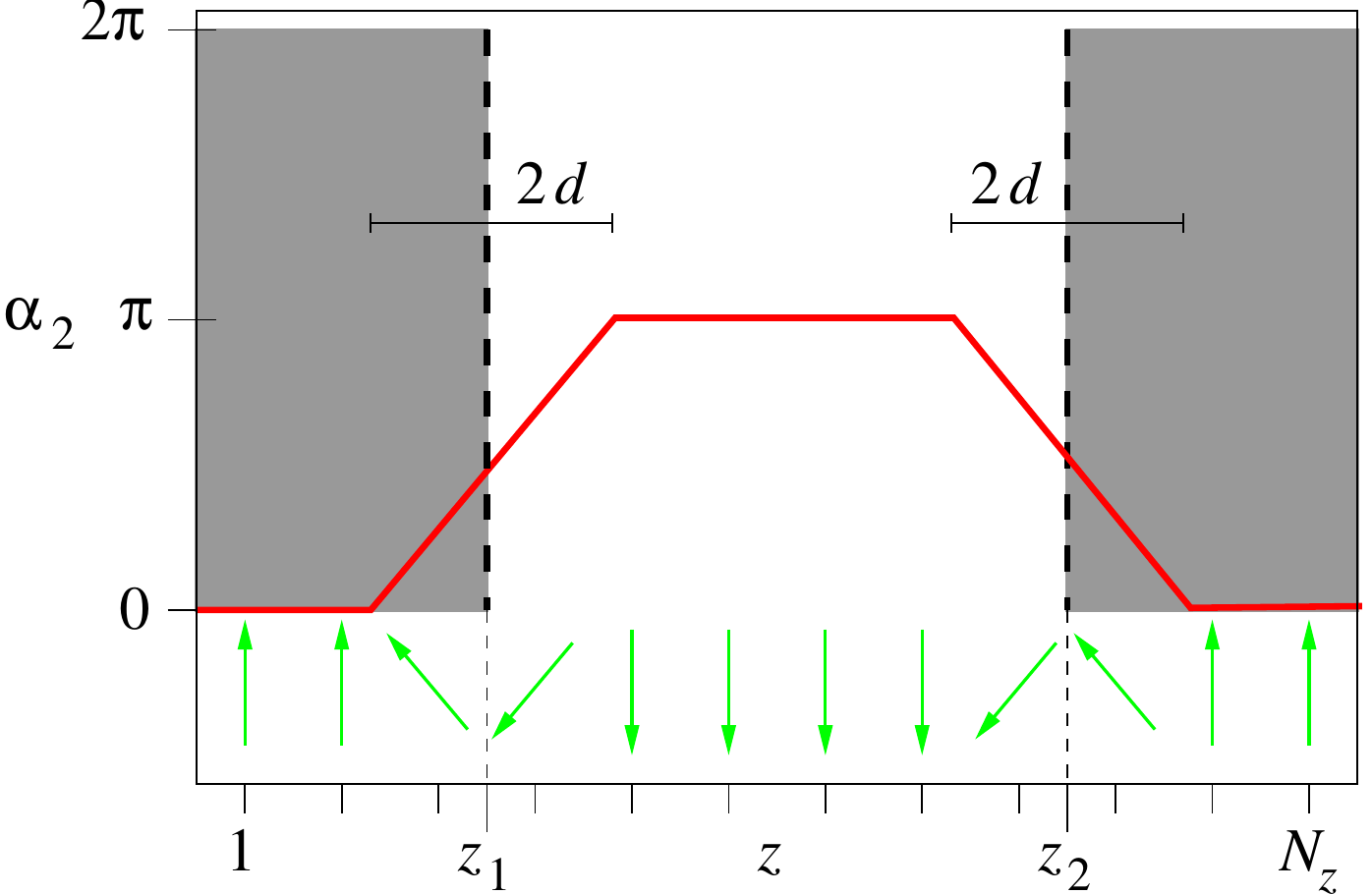}
\caption{The link angle a) $\alpha_1$ of a parallel and b) $\alpha_2$ of an anti-parallel $xy$ vortex pair. The arrows ($t$-links) rotate counterclockwise with increasing $\alpha_i$ in $z$ direction. The vertical dashed lines indicate the positions of vortices after center projection. In the shaded areas the links have positive, otherwise negative trace.}\label{fig:phis}
\end{figure}
These profile functions are plotted in Fig.~\ref{fig:phis}. As shown, upon traversing a vortex sheet, the angle $\alpha$ increases or decreases by
$\pi$ within a finite thickness $2d$ of the vortex. The vortex
pairs with the same (opposite) vortex orientation are called parallel (anti-parallel) vortices.

Unicolor plane vortices can contribute to the topological charge density through intersections. As shown in~\cite{Engelhardt:1999xw}, each intersection between two unicolor vortex sheets carries a topological charge with modulus $|Q|=1/2$, whose sign depends on the relative orientation of the vortex fluxes.

\begin{figure}[h!]
\centering
\includegraphics[scale=0.8]{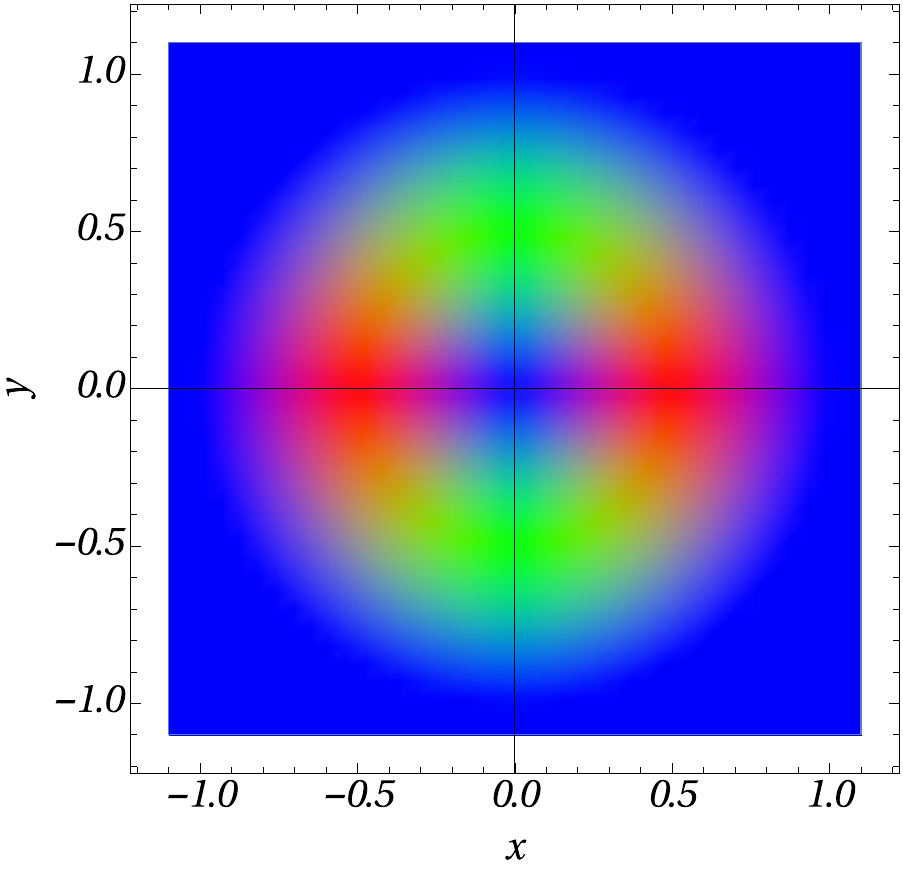}
\caption{The color direction $\vec n$ of Eq.~(\ref{Dirn}) in the  xy-plane for $R=1$ and $x_0=y_0=0$ by maps to RGB-colors, $\pm\hat i\mapsto$ red, $\pm\hat j\mapsto$ green and $\pm\hat k\mapsto$ blue.}
\label{fig:cylinder}
\end{figure}

Now, we introduce a color structure for one of the plane vortices by the links
\begin{equation}\label{originallinks}
U_i(x)=\mathbf 1,\quad U_4(x)=
\begin{cases}U^\prime_4(\vec x)&\mathrm{for}\quad t=1,\\\mathbf 1&\mathrm{else},
\end{cases}
\end{equation}
where
\begin{equation}\label{eq:sphv}
U^\prime_4(\vec x)=
\begin{cases}
\mathrm e^{\mathrm i\alpha(z)\vec n\cdot\vec\sigma}\quad&\mathrm{for}\quad
z_1-d\leq z\leq z_1+d\quad\mathrm{and}\quad 0\leq\rho\leq R,
\\\mathrm e^{\mathrm i \alpha(z)\sigma_3}  & \mathrm{else}.\end{cases}
\end{equation}
The color direction $\vec n$ in $U^\prime_4(\vec x)$ is defined by
\begin{equation}\label{Dirn}
\vec n=\hat i\,\sin\theta(\rho)\cos\phi+\hat j\,\sin\theta(\rho)\sin\phi
+\hat k\,\cos\theta(\rho),
\end{equation}
where
\begin{equation}
\rho=\sqrt{(x-x_0)^2+(y-y_0)^2},\quad
\theta(\rho)=\pi(1-\frac{\rho}{R}),
\end{equation}
and
\begin{equation}
\phi=\arctan_2\frac{y-y_0}{x-x_0}\;\in\,[0,2\pi).
\end{equation}
In Fig.~\ref{fig:cylinder} the color direction $\vec n$ is displayed in the xy-plane for $R=1$ and $x_0=y_0=0$ by maps to RGB-colors, $\pm\hat i\mapsto$ red, $\pm\hat j\mapsto$ green and $\pm\hat k\mapsto$ blue.

The colorful region is located in the range
\begin{equation}\label{ColCylReg}
0 \leq \rho \leq R\quad\textrm{and}\quad z_1-d\leq z \leq z_1+d,\quad
\textrm{the ``colorful cylindrical region''}.
\end{equation}
In this cylindrical region with the center at $(x_0,y_0)$ only the temporal links $U_4$ are nontrivial and all spatial links $U_i$ are trivial. Therefore, for this configuration the gluonic lattice topological charge $Q$ is zero. In order to show that these vortices define a vacuum to vacuum transition we apply a gauge transformation to the lattice links given in Eq.~(\ref{originallinks}). The lattice gauge transformation is considered \cite{Schweigler:2012ae}
\begin{equation}
\Omega(x)=\begin{cases}
g(\vec{x})&\mathrm{for}\quad1<t\leq t_g,\\\mathbf 1&\mathrm{else},\end{cases}
\end{equation}
where
\begin{equation}\label{g}
g(\vec{x})=[U^\prime_4(\vec x)]^\dagger,
\end{equation}
Therefore, the links of the plane vortices become \cite{Schweigler:2012ae}
\begin{align}\begin{split}\label{vvlattoriginal}
U_i(x)&=\begin{cases}
g(\vec x+\hat i)\,g(\vec x)^\dagger\quad&\mathrm{for}\quad1< t\leq t_g,\\
\mathbf 1&\mathrm{else},\end{cases}\\U_4(x)&=\begin{cases}
g(\vec x)^\dagger\quad&\mathrm{for}\quad t=t_g,\\\mathbf 1&\mathrm{else}.
\end{cases}
\end{split}\end{align}
From this it becomes clear that plane vortices represent a
transition in the temporal direction between two pure gauge fields. The
transition occurs between $t=1$ and $t=2$. The winding number is defined by
\begin{gather}\label{eq:winding}
N_w=-\frac{1}{24\pi^2}\int\mathrm d^3x\,\epsilon_{ijk}\,
\mathrm{Tr}[g^\dag\partial_i g\;g^\dag\partial_jg\;g^\dag\partial_k g].
\end{gather}
For $t\leq1$, where we have the trivial gauge field we get $N_{w1}=0$. For $t>1$ the pure gauge field is generated by the gauge transformation in Eq.~(\ref{g}) with winding number $N_{w2}=-1$. Therefore in the continuum limit, colorful plane vortices have topological charge $Q=N_{w2}-N_{w1}=-1$.

Assuming an infinitely big temporal extent of the lattice and taking $t_g\to\infty$, the continuum field corresponding to Eq.~(\ref{vvlattoriginal})
can be written as
\begin{equation}\label{vvtransgeneral}
\mathcal A_\mu=\mathrm i\,f(t)\,\partial_\mu g\,g^\dagger,
\end{equation}
where $g$ is the gauge transformation given in Eq.~(\ref{g}) and $f(t)$
determining the transition in temporal direction $t$, is a step function.
Clearly, one could use a smoother function $f(t)$ that changes more slowly than
a sudden jump between $0$ and $1$. We refer to plane vortices with a smoother
function $f(t)$ as generalized plane vortices. Their lattice version will be
derived from the continuum form in section \ref{sec:gen_latt}. Before doing
that, the action and topological charge density of the continuum gauge field are investigated in the next two sections.

\section{Action of the plane vortices in the continuum}
\label{sec:action}
Now, we calculate the continuum action $S$ for the generalized plane vortices in
Euclidean space with $\mathcal A_\mu$ of Eq.~(\ref{vvtransgeneral}). We insert $g$ of Eq.~(\ref{g}) and write
\begin{equation}
g=\begin{cases}q_0\,\sigma_0+\mathrm i\vec q\cdot\vec\sigma
&\textrm{for colorful, cylindrical region},\\
k_0\,\sigma_0+\mathrm i\vec k\cdot\vec\sigma&\mathrm{else},\end{cases}
\end{equation}
with $q_0=k_0=\cos\alpha(z)$, $\vec q=-\vec n\,\sin\alpha(z)$, and $\vec k=-\hat k\,\sin\alpha(z)$. We get
\begin{equation}\label{genpqqd}
\mathrm i\,\partial_\mu g\,g^\dagger=
\begin{cases}
\vec\sigma\cdot[\partial_\mu q_0\,\vec q-q_0\,\partial_\mu\vec q+
\vec q\times\partial_\mu\vec q]&\textrm{for colorful cylindrical region},\\
\vec\sigma\cdot[\partial_\mu k_0\,\vec k-   k_0\,\partial_\mu\vec k+
\vec k\times\partial_\mu\vec k\,]&\mathrm{else}.
\end{cases}
\end{equation}
Inserting this into
\begin{equation}
\mathcal{A}_{\mu} =
\begin{cases}
\frac{\sigma^a}{2}A_\mu^{1a}&\textrm{for colorful, cylindrical region},\\
\frac{\sigma^a}{2} A_\mu^{2a}&\mathrm{else},\
\end{cases}
\end{equation}
gives after a few lines of calculations
\begin{align}\begin{split}\label{a_log}
&A_i^{1a}=2f(t)\left[\alpha'(z)\,n_a\delta_{i3}
        +\cos\alpha(z)\sin\alpha(z)\,\partial_in_a
+\sin^2\alpha(z)\,(n_l\partial_in_m\epsilon_{lma})\right],\\
&A_{4}^{1a}=0,
\end{split}\end{align}
and outside the colorful, cylindrical region, including the unicolor sheet
\begin{align}\begin{split}\label{b_log}
&A_i^{2a}=2f(t)\,\alpha^\prime(z)\,\delta_{a3}\,\delta_{i3},\\
&A_4^{2a}=0.
\end{split}\end{align}
In the following, we use the notation
\begin{equation}
A_\mu^a=f(t)\,A_{\mu+}^a\quad\textrm{with}\quad
A_{\mu+}^a=\mathrm i\,\partial_\mu g\,g^\dagger,
\end{equation}

Using the antisymmetry property of $\mathcal{F}_{\mu\nu}$ we obtain
\begin{align}\begin{split}\label{trsplit}
\mathrm{tr}_C\left[ \mathcal{F}_{\mu \nu}\mathcal{F}_{\mu \nu}\right]
&=\mathrm{tr}_C\left[\mathcal{F}_{ij}\mathcal{F}_{ij}\right]
+2\,\mathrm{tr}_C\left[\mathcal{F}_{4i}\mathcal{F}_{4i}\right] ,
\end{split}\end{align}
where the field strength tensor is given by $\mathcal{F}_{\mu\nu}=\frac{\sigma^j}{2}F_{\mu \nu}^j$ with $F_{\mu \nu}^j=\partial_\mu A_\nu^j-\partial_\nu A_\mu^j-\epsilon_{jkl}A_\mu^kA_\nu^l$. Since the gauge field has no temporal components, {\it i.e.} $A_4^a=0$, we can simplify $F_{4i}^a$ to
\begin{equation}\label{elsimpl}
F_{4i}^a=\partial_4A_i^a.
\end{equation}
With Eq.~(\ref{elsimpl}) we get \cite{Schweigler:2012ae}
\begin{equation}
\mathrm{tr}_C\left[\mathcal{F}_{4i}\mathcal{F}_{4i}\right]=\frac{1}{2}\left(\frac{d}{d t}f (t)\right)^2A_{i+}^aA_{i+}^a.
\end{equation}
The $F_{i j}^a$ can be simplified to \cite{Schweigler:2012ae}
\begin{align}\label{magsimpl}
F_{ij}^a&=f(t)[1-f(t)]\,\epsilon^{abc}A_{i+}^bA_{j+}^c.
\end{align}
Inserting the explicit form of $A_{i+}^a$ we obtain
\begin{equation}\label{a}
A_{i+}^aA_{i+}^a=
\begin{cases}
4\left\{\alpha'(z)^2+\frac{1}{{\rho}^2}\sin^2\alpha(z)
\left[\sin^2\theta(\rho)+\rho^2\theta^\prime(\rho)^2\right]\right\}
&\textrm{for colorful cylindrical region},\\
4\,\alpha^\prime(z)^2&\mathrm{else},
\end{cases}
\end{equation}
and with Eq.~(\ref{magsimpl}) \cite{Schweigler:2012ae}
\begin{equation}
\mathrm{tr}_C\left[\mathcal{F}_{ij}\mathcal{F}_{ij}\right]
=\frac{1}{2}f(t)^2[1-f(t)]^2\,\epsilon^{abc}A_{i+}^b A_{j+}^c\,
   \epsilon^{ade}A_{i+}^dA_{j+}^e.
\end{equation}
Using a computer algebra program we get for the colorful cylindrical region
\begin{equation}\begin{aligned}
\epsilon^{a b c}&A_{i+}^bA_{j+}^c\,\epsilon^{ade}A_{i+}^dA_{j+}^e=\\
&=\frac{32}{\rho^2}\sin^2\alpha(z)
\left[\sin^2\alpha(z)\sin^2\theta(\rho)\,\theta^\prime(\rho)^2
+\alpha'(z)^2\Big(\sin^2\theta(\rho)+\rho^2\theta^\prime(\rho)^2\Big)\right],
\end{aligned}\end{equation}
and $0$ outside of this region.
We will choose $f(t)$ as the piecewise linear function
\begin{equation}\label{fk_ramp}
f_{\Delta t}(t)=\begin{cases}0&\mathrm{for}\quad t<1,\\\frac{t-1}{\Delta t}
&\mathrm{for}\quad 1\leq t\leq1+\Delta t,\\1
&\mathrm{for}\quad t>1+\Delta t,\end{cases}
\end{equation}
where $\Delta t$ stands for the duration of the transition.

Switching to cylindrical coordinates, we get the action
\begin{align}\begin{split}\label{sktempdone}
S^1=\frac{1}{2g^2}\Bigg\{&\Delta t\int\mathrm d\rho\int\mathrm dz\,
\frac{16\pi}{15\rho}\sin^2\alpha(z)\times\\
&\times\left[\sin^2\alpha(z)\sin^2\theta(\rho)
\,\theta^\prime(\rho)^2+\alpha'(z)^2
\Big(\sin^2\theta(\rho)+\rho^2\theta^\prime(\rho)^2\Big)\right]+\\
+&\frac{1}{\Delta t}\int\mathrm d\rho\int\mathrm dz\,8\pi
\left[\rho\,\alpha'(z)^2+\frac{1}{\rho}\sin^2\alpha(z)\Big(\sin^2\theta(\rho)
+\rho^2\theta^\prime(\rho)^2\Big)\right]\Bigg\},
\end{split}\end{align}
for the colorful cylindrical region~(\ref{ColCylReg}) and the action
\begin{align}\begin{split}\label{sktempdon}
S^2=\frac{1}{2 g^2}\frac{1}{\Delta t}
\int{\mathrm d\rho\int{\mathrm dz\,8\pi\,\rho\,\alpha^\prime(z)^2}}
\end{split}\end{align}
for the ``unicolor cylindrical region''
\begin{equation}
0 \leq \rho \leq R\quad\textrm{and}\quad z_2-d\leq z \leq z_2+d.
\end{equation}
As long as none of the spatial integrations gives zero, the action $S^1$
diverges for both $\Delta t\to\infty$ and $\Delta t\to0$ and the action $S^2$
diverges for $\Delta t\to0$ and converges to zero for $\Delta t\to\infty$. The
action has electric and magnetic contributions in the colorful region but in the
unicolor region the magnetic term vanishes. Therefore, only the colorful region of plane vortices contributes to the topological charge.

\begin{figure}[h!]
\centering
a)\includegraphics[width=0.46\columnwidth]{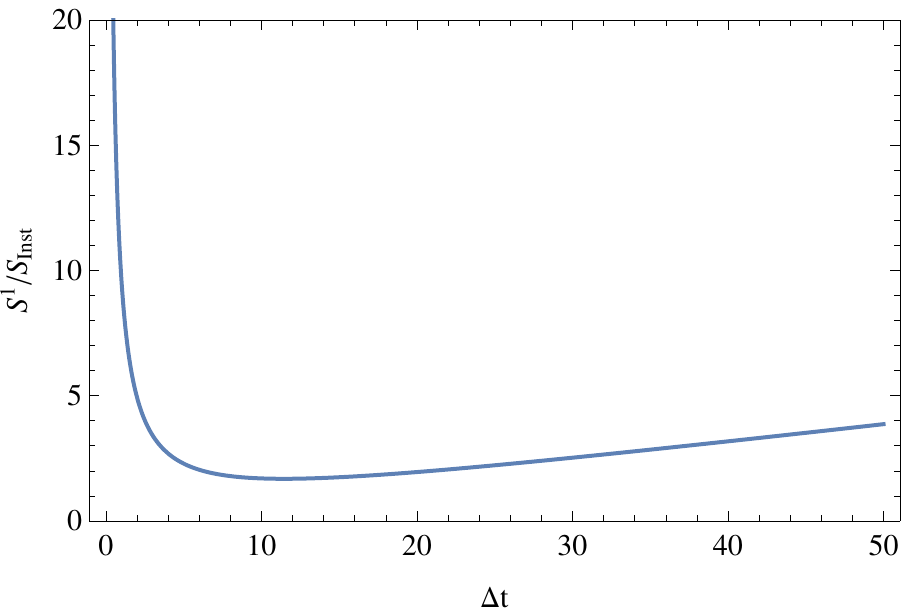}
b)\includegraphics[width=0.46\columnwidth]{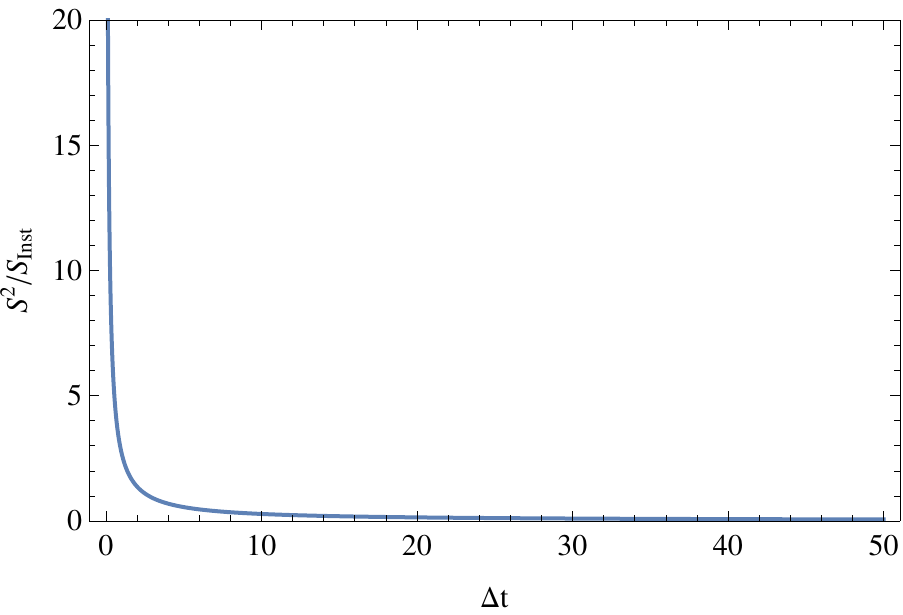}
\caption{The gauge action in units of the instanton action $S_\mathrm{Inst}$ a)
for the colorful region $0 \leq \rho \leq R$ and $z_1-d \leq z \leq z_1+d$ where
both electric and magnetic terms contribute. The minimum of the action is
$S^1_\mathrm{min}=1.68\,S_\mathrm{Inst}$. b) Action for the unicolor region $0 \leq \rho \leq R$ and $z_2-d \leq z \leq z_2+d$, with an electric term only.}
\label{fig:action1}
\end{figure}

Performing the spatial integration in Eqs. (\ref{sktempdone}) and (\ref{sktempdon}) and using $\alpha_2(z)$ defined in Eq.~(\ref{eq:phi-pl0}) we get at $d/R=1$ with a computer algebra program
\begin{equation}
S=\begin{cases}
\frac{S^1(\Delta t)}{S_\mathrm{Inst}}=\frac{0.51\,\Delta t}{R}
+\frac{1.37\,R}{\Delta t}
&\textrm{for colorful cylindrical region},\\
\frac{S^2(\Delta t)}{S_\mathrm{Inst}}=\frac{0.39\,R}{\Delta t}&\textrm{for unicolor cylindrical region}.
\end{cases}
\end{equation}
The gauge actions for the colorful and unicolor regions as functions of the temporal extent $\Delta t$ for $R=d=7$ on a $28^3\times40$ lattice are plotted in Fig.~\ref{fig:action1} a) and b) respectively. The gauge actions are in units of the instanton action $S_\mathrm{Inst}=8\pi^2/g^2$. This action serves as a lower bound for objects with topological charge $|Q| = 1$. For slow transitions, the action of the unicolor region approaches zero and in the colorful region the action is finite due to both electric and magnetic terms.

\section{Topological charge of plane vortices in the continuum}
\label{sec:topdencont}
In this section, we calculate the topological charge in the continuous Euclidean space for the generalized plane vortices. The topological charge density $q(x)$ can be written as a total derivative \cite{Diakonov:2002fq}
\begin{equation}\label{q}
q(x)=\partial_\mu K_\mu(x)\quad\mathrm{with}\quad
K_\mu=-\frac{1}{16\pi^2}\epsilon_{\mu\alpha\beta\gamma}
\left(A_\alpha^a \partial_\beta A_\gamma^a
-\frac{1}{3}\epsilon^{abc}A_\alpha^a A_\beta^b A_\gamma^c\right).
\end{equation}
We again use the shorthand notation $\mathcal A_{i+}=\mathrm i\,\partial_i g\,g^\dagger$. The gauge field
in Eq. (\ref{vvtransgeneral}) has no temporal component, {\it i.e.} $\mathcal{A}_4 = 0$. Therefore, the spatial component $K_i$ is zero. Using Eq.~(\ref{q}) the temporal component $K_4$ reads \cite{Schweigler:2012ae}
\begin{equation}\label{k4simp}
K_4=\frac{1}{16\pi^2}\left[\left(\frac{1}{2}f(t)^2
-\frac{1}{3}f(t)^3\right)\epsilon_{ijk}\,
\epsilon^{abc}A_{i+}^aA_{j+}^bA_{k+}^c\right].
\end{equation}
As a result, the topological charge density $q(x)$ is the temporal derivative of the temporal component $K_4$. Using $g$ defined in Eq.~(\ref{g}), $\alpha_2(z)$ from Eq.~(\ref{eq:phi-pl0}) and $f(t)$ in Eq.~(\ref{fk_ramp}) the topological charge density is
\begin{equation}
q(\rho,z,t)=
\begin{cases}
q^\prime(\rho,z,t)\quad&\mathrm{for}\quad0\leq\rho\leq R,\quad
z_1-d\leq z\leq z_1+d\quad\mathrm{and}\quad1\leq t\leq1+\Delta t,\\
0&\mathrm{else},
\end{cases}
\end{equation}
where
\begin{equation}\label{q^1}
q^\prime(\rho,z,t)=-\frac{3\pi}{\rho\,Rd}\sin\left(\pi(1-\frac{\rho}{R})\right)
\sin^2\left(\frac{\pi}{2d}(d+z-z_1))\right)\left(\frac{1}{4\Delta t}
-\frac{(t-1-\frac{\Delta t}{2})^2}{\Delta t^3}\right).
\end{equation}
Integrating over $\rho\,\mathrm d\rho\,\mathrm dz$ and $\mathrm dt$ we get the topological charge
\begin{equation}
Q=\begin{cases}
-1&\quad\textrm{for the colorful region},\\
0&\quad\textrm{for the unicolor region}.
\end{cases}
\end{equation}

\section{The generalized plane vortices on the lattice}\label{sec:gen_latt}
Now, the generalized continuum plane vortices are put on the lattice with
periodic boundary conditions. $\mathcal{A}_\mu$, as defined in Eq.~(\ref{vvtransgeneral}), vanishes for $t\to-\infty$ but not for $t\to\infty$. Therefore, this field configuration does not fulfill periodic boundary conditions in the temporal direction. For getting a vanishing gauge field $\mathcal A_\mu$ for $t\to\infty$, one can use a lattice gauge transformation that equals $\mathbf 1$ for $t\to-\infty$ and $g^\dagger$ for $t\to\infty$. Therefore the lattice links for the smoothed continuum plane vortices with $f(t)$ given in Eq.~(\ref{fk_ramp}) are
\begin{align}\begin{split}\label{reallattobj}
U_{i}(x)&=\begin{cases}
\left[g(\vec r+\hat i)\,g(\vec r)^\dagger\right]^{(t-1)/\Delta t}
& \mathrm{for}\quad1<t<1+\Delta t,\\
g(\vec{r}+\hat i)\,g\left(\vec r\right)^{\dagger}
& \mathrm{for}\quad1+\Delta t\leq t\leq t_g,\\
\mathbf 1&\mathrm{else},
\end{cases}\\
U_4(x)&=\begin{cases}
g(\vec r)^\dagger&\mathrm{for}\quad t=t_g,\\
 \mathbf 1&\mathrm{else},
\end{cases}
\end{split}\end{align}
where the functions $g(\vec r)$ and $\alpha(z)$ are defined in Eqs.~(\ref{g}) and (\ref{eq:phi-pl0}) respectively.

\begin{figure}[h!]
\centering
a)\includegraphics[width=0.46\columnwidth]{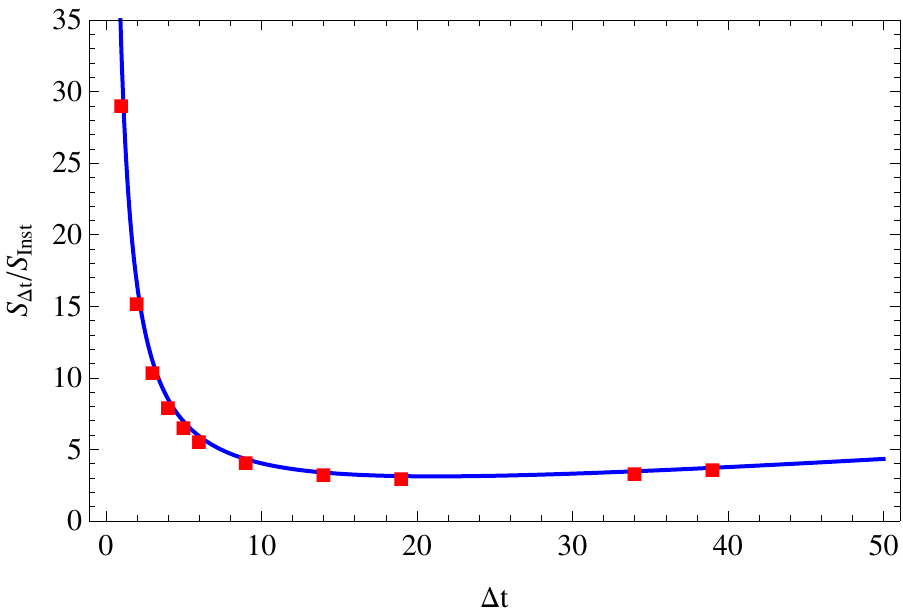}
b)\includegraphics[width=0.46\columnwidth]{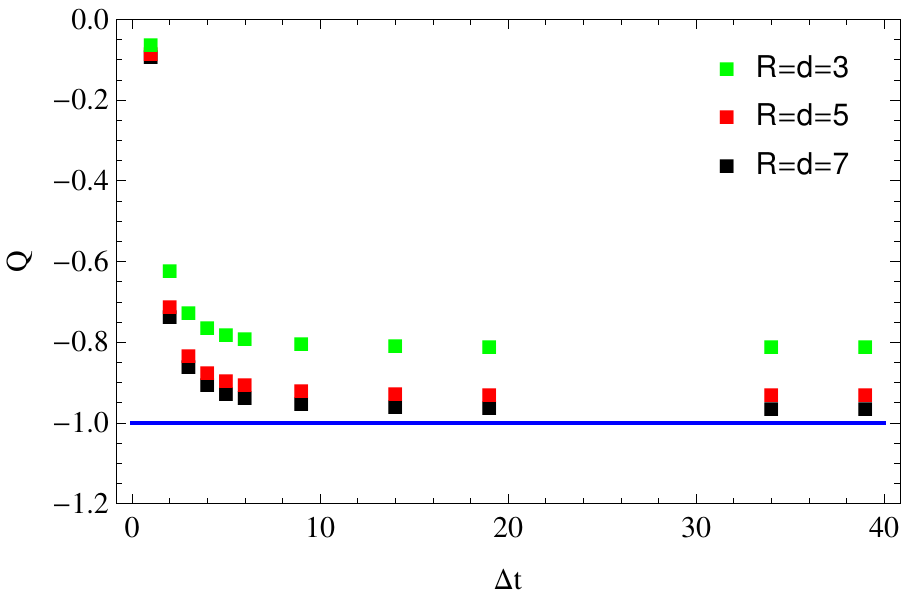}
\caption{a) Comparison of the gauge action of plane vortices in units of the instanton action $S_\mathrm{Inst}$ on the lattice (squares) with the one in the continuum (line) as a function of the temporal extent $\Delta t$ of the plane vortices. The radius and the thickness of the vortex are given by $R = 7$ and $d = 7$. The minimum of the action is $S_\mathrm{min}=3.1\,S_\mathrm{Inst}$. b) Comparison of the gluonic topological charge $Q(\Delta t)$ on the lattice for three
values of $R=d=3,5,7$ (calculated with the plaquette definition) with the one in
the continuum. In the continuum we get $Q(\Delta t) = -1$ for all $\Delta t$ as
shown by the horizontal line. On the lattice, the topological charge for slow
transitions increasing the radius and thickness of the plane vortices converges
to near $-1$. The calculations have been performed on $28^3 \times 40$ lattices with lattice constant $a=1$.}\label{fig:s_latt_vs_ana}
\end{figure}
In Fig.~\ref{fig:s_latt_vs_ana} we plot the action and topological charge of this lattice configuration for a $28^3 \times 40$ lattice. As shown in  Fig.~\ref{fig:s_latt_vs_ana}a, the lattice action matches the continuum action very well. The topological charge as a function of $\Delta t$ for three values of $R=d$ is plotted in Fig.~\ref{fig:s_latt_vs_ana}b). At $\Delta t = 1$, the fast vacuum to vacuum transition, the topological charge is close to zero. This is a lattice artifact. The continuum value of the topological charge, $Q=-1$, is approached for reasonably large values of $\Delta t$, $R$ and $d$.

\begin{figure}[h!]
\centering
\includegraphics[width=0.47\columnwidth]{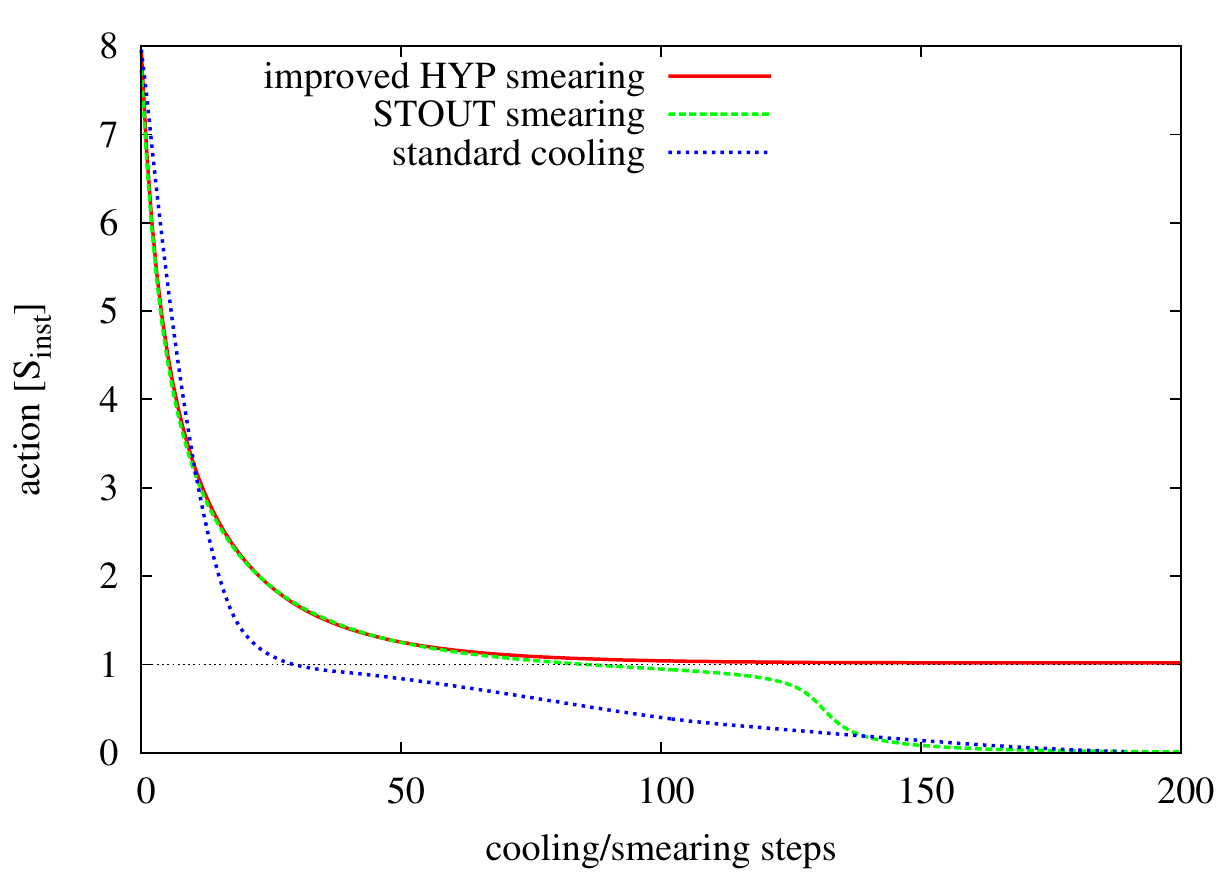}
\includegraphics[width=0.47\columnwidth]{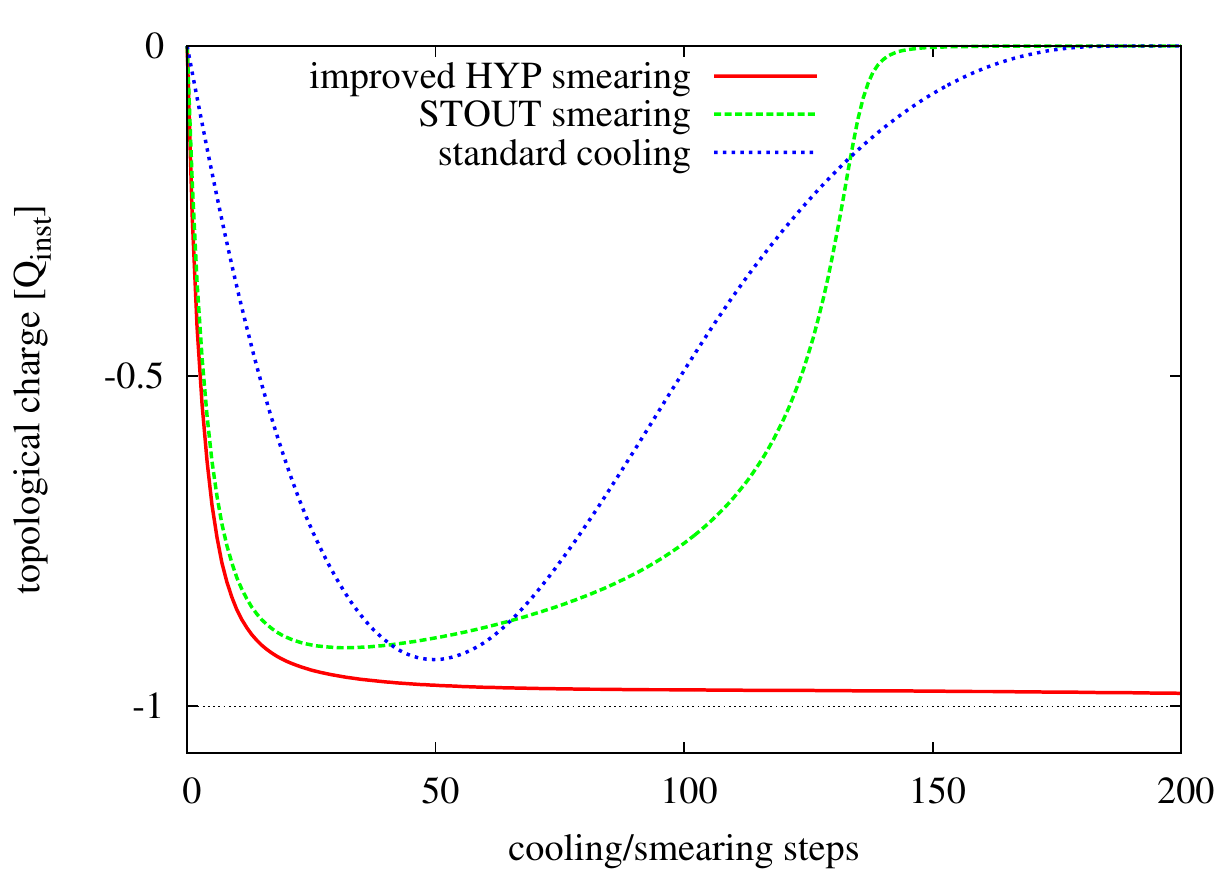}
\includegraphics[width=0.47\columnwidth]{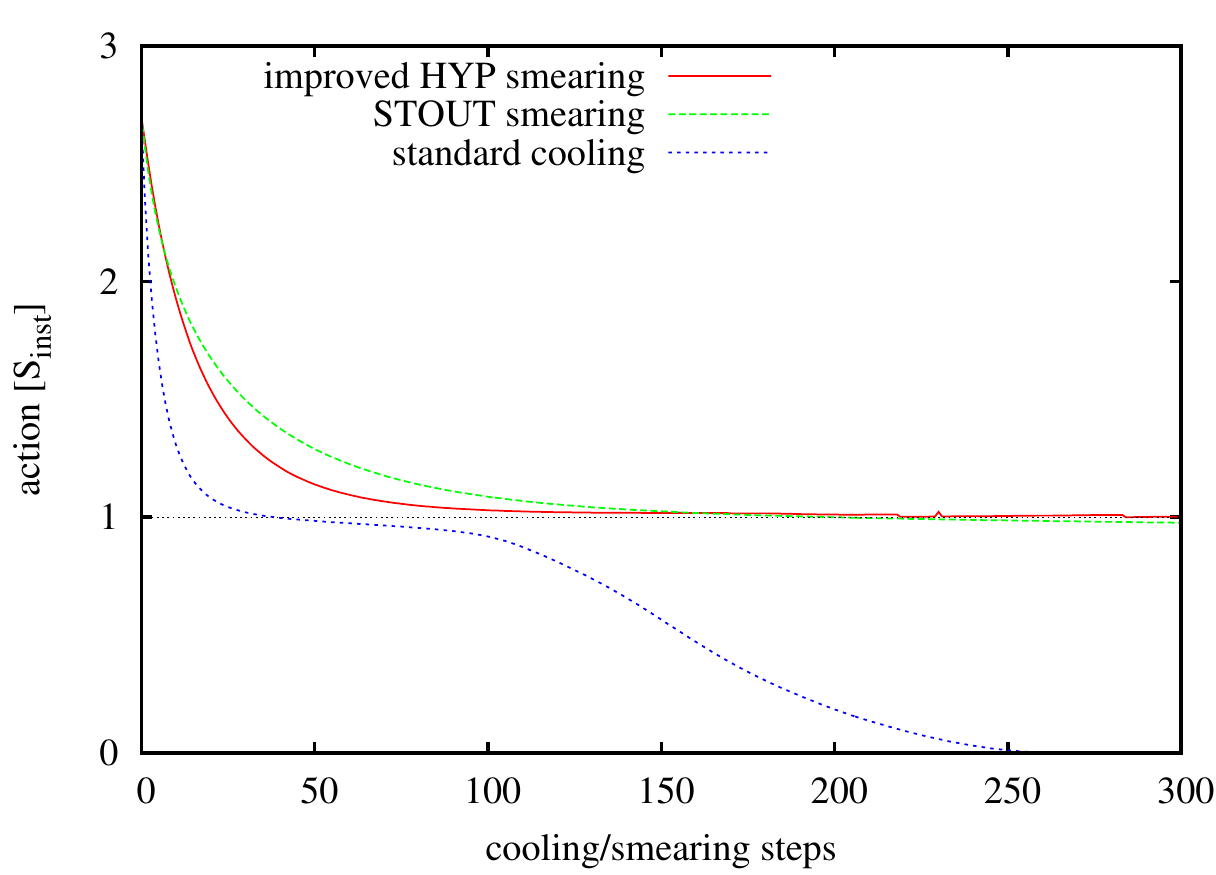}
\includegraphics[width=0.47\columnwidth]{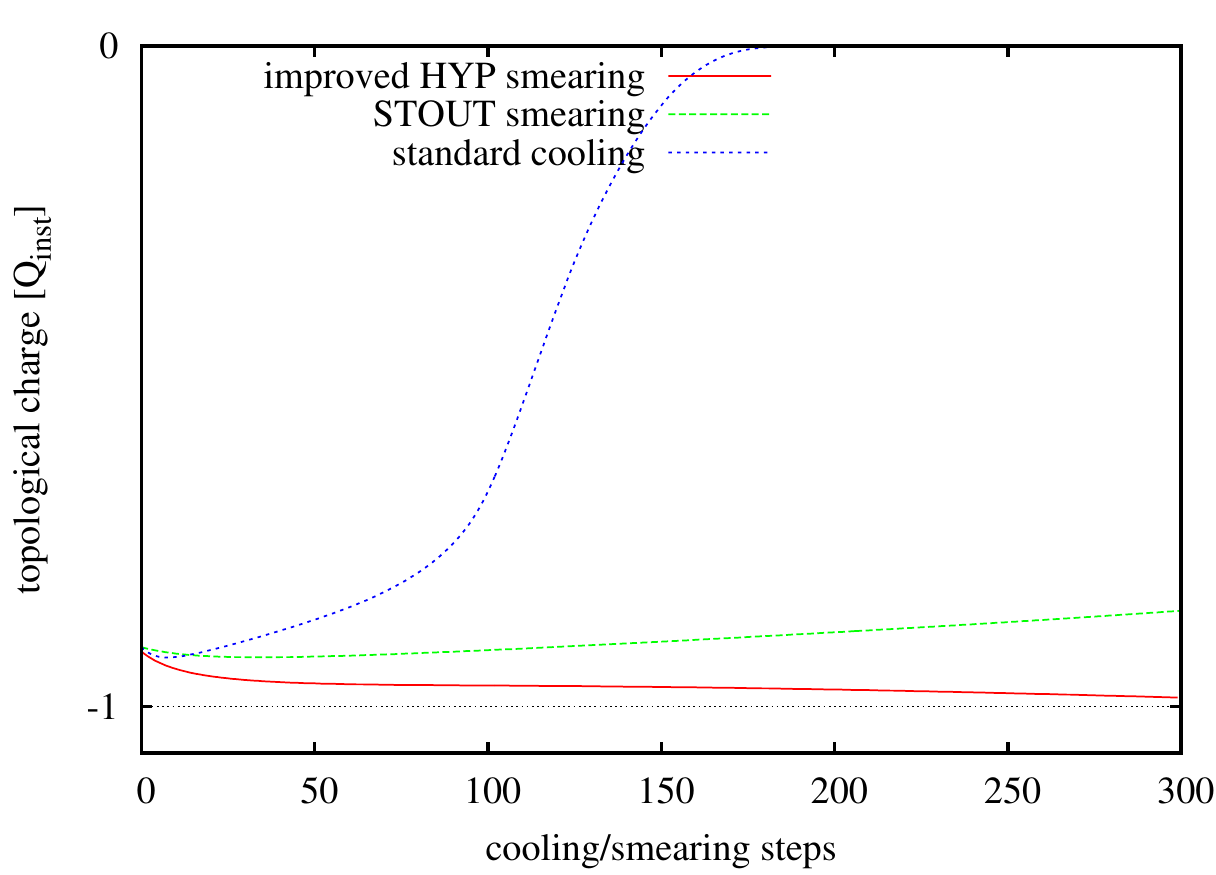}
\caption{Cooling, STOUT and improved HYP smearing histories of total
	action and topological charge of the generalized colorful plane vortex
	configuration with $\Delta t = 1$ (above) and $\Delta t = 11$ (below).
	As expected we recover one instanton action and (negative) topological charge, the
	smoothed configuration (below) immediately gives the correct topological
	charge and lower action, it also seems to be more stable under STOUT smearing. Only the improved smearing version however stabilizes the topological object leading to a perfect spherical symmetric anti-instanton.
}\label{fig:actop}
\end{figure}

As expected, cooling or smoothing of these generalized vortices leads to
anti-instanton configurations with four dimensional spherical symmetry and their
center located in the center of the rotational symmetric color structure of the
colorful plane vortex shown in Fig.~\ref{fig:cylinder}. Actually, with standard
cooling and simple STOUT smearing the topological object falls through the
lattice before developing the perfect spherical symmetry, improved HYP
smearing however stabilizes the topological charge contribution and reveals the
anti-instanton. Cooling and various smearing histories of the total action and
topological charge are shown in Fig.~\ref{fig:actop} for $\Delta t = 1$ (above) and $\Delta t = 11$ (below). The results are in
accordance with~\cite{Trewartha:2015ida} where it was shown that vortex-only
configurations, {\it i.e.} the center vortex content projected out of full
$SU(3)$ Monte Carlo configurations after Maximal Center Gauge, reveal the
instanton--anti-instanton content after gradient flow or over-improved cooling,
showing that the long-range structure is contained within the center vortex
degrees of freedom.

\section{Dirac eigenmodes for plane vortices}\label{sec:Dirac}
In QCD a non zero value of the chiral condensate $\left\langle \bar{\psi}\psi \right\rangle$, the chiral order parameter, indicates spontaneous chiral symmetry breaking. According to the Banks-Casher analysis \cite{Banks:1979yr}, a finite density of near-zero eigenmodes of the chiral-invariant Dirac operator leads to a
finite chiral condensate. Therefore, for studying the effect of colorful plane
vortices on fermions, we determine the low-lying eigenvectors and eigenvalues
$|\lambda| \in [0,1]$ of the overlap Dirac operator $D_\mathrm{ov}$
\cite{Hollwieser:2013xja} as a Ginsparg-Wilson operator. The absolute value
$|\lambda|$ of the two complex conjugate eigenvalues $\lambda$ and $\lambda^*$
(doubler modes) is simply written as $\lambda$ if $\lambda\notin\{0,1\}$. Therefore, for every $\lambda\notin\{0,1\}$, we have two eigenvectors $\psi_\pm$, with equal scalar and chiral densities. For convenience, we enumerate the eigenmodes in ascending order of the eigenvalues. \#0+ means a right-handed zero mode, \#0-- a left-handed zero mode and \#1  the lowest nonzero mode etc. Their eigenvalues are referred to as $\lambda$\#0+, $\lambda$\#0--, $\lambda$\#1, etc., and their densities as $\rho$\#0+, etc.

The chiral density of the eigenvectors $\psi_\pm$ which is important to assess the local chirality properties is given by \cite{Hollwieser:2013xja}

\begin{equation}\label{eq:densities}
\rho_5=\psi^\dagger_\pm\gamma_5\psi_\pm =\rho_+-\rho_-,
\end{equation}
where $\rho_+$ and $\rho_-$ are left- and right-handed chiral densities. According to the Atiyah-Singer index theorem, the topological charge is given by the index
\begin{equation}\label{eq:index}
\mathrm{ind}D[A]=n_--n_+=Q,
\end{equation}
 where  $n_-$ and $n_+$ denote the numbers of left- and right-handed zero modes
\cite{Atiyah:1971rm,Schwarz:1977az,Brown:1977bj}
. We remark that one never finds zero modes of both chiralities for a single configuration using usual antiperiodic boundary conditions in time direction.
Therefore in a gauge field with topological charge $Q\neq 0$, $D_\mathrm{ov}$ has $|Q|$ exact zero modes with chirality $-\mathrm{sign} Q$.
\begin{figure}[h!]
\centering
a)\includegraphics[width=0.47\columnwidth]{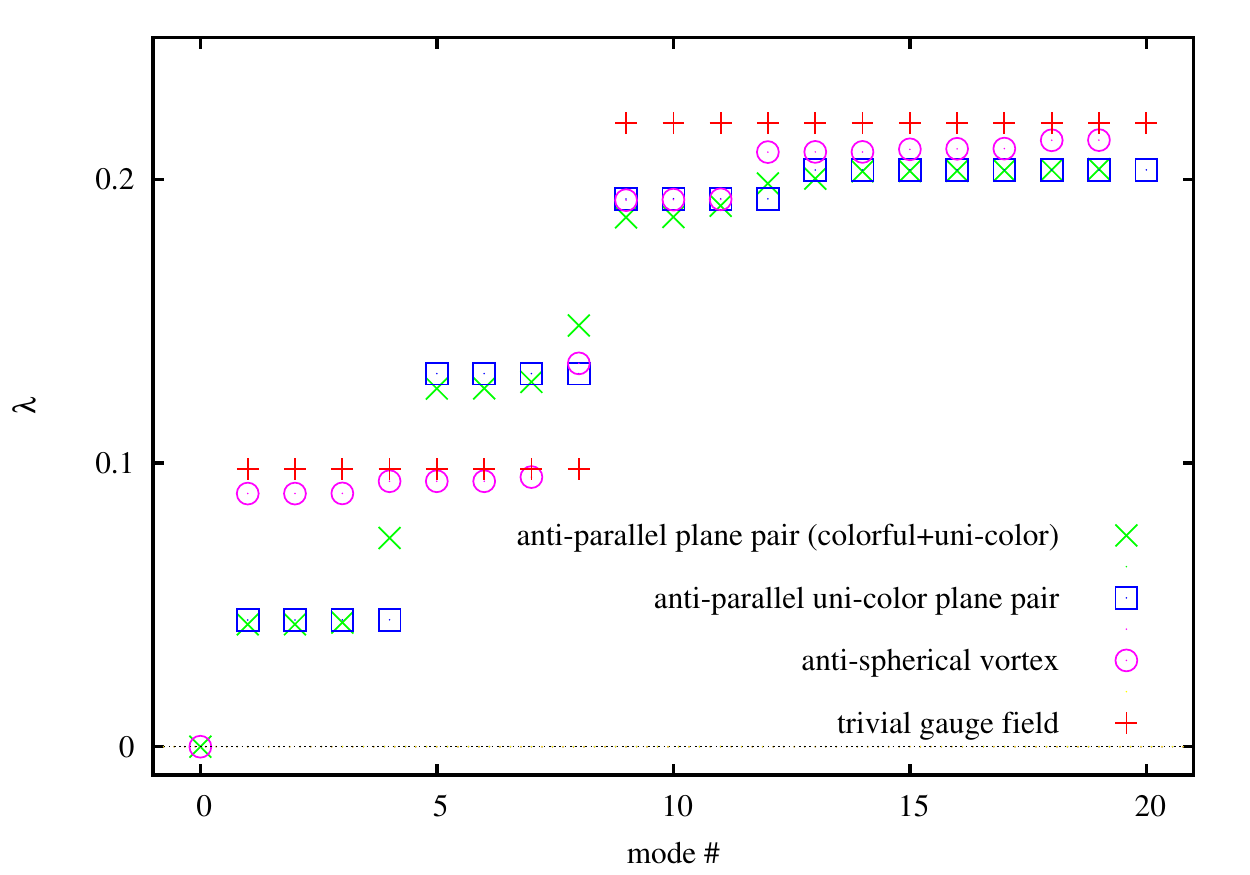}
b)\includegraphics[width=0.47\columnwidth]{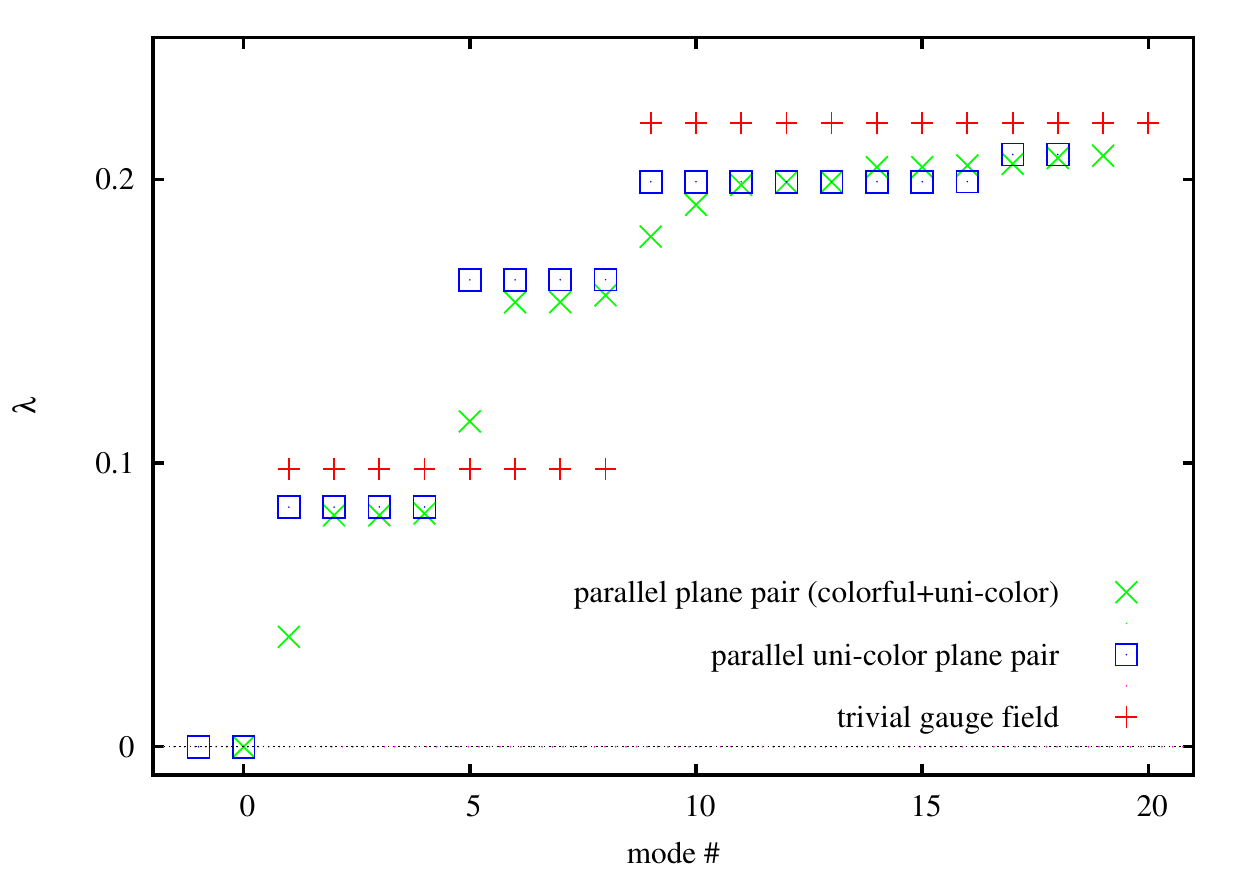}
\caption{The lowest overlap eigenvalues a) for anti-parallel colorful plane vortices, anti-parallel unicolor plane vortices, and an (``anti''-)spherical vortex with $Q=-1$ compared to the eigenvalues of the free (overlap) Dirac operator. b) for parallel colorful plane vortices and parallel unicolor plane vortices on a $16^4$ lattice. In the figures "colorful+unicolor" means one sheet of the plane vortex pair is colorful and the other one is unicolor.}
\label{overlap}
\end{figure}
In Fig.~\ref{overlap}, we show the lowest overlap eigenvalues for colorful plane vortices, unicolor plane vortices, and a (``anti''-)spherical vortex with $Q=-1$ compared to the eigenvalues of the free overlap Dirac operator on a $16^4$-lattice. For fermionic fields we use periodic boundary conditions in spatial and anti-periodic in temporal directions. For colorful plane vortices, the colorful and unicolor vortex sheets with thickness $d=3$ are located around $z_1=4.5$ and $z_2=12.5$ respectively. The center of the colorful region with radius $R = 8$ is located in the $xy$ plane around $x=y=8$. The parameters for unicolor plane vortices are the same as for the colorful plane vortices. For the spherical vortices, using the ansatz $\alpha_+$ for the profile function $\alpha$ in Ref. \cite{Hollwieser:2013xja}, the center of the configuration with core radius $R = 5.5$ and thickness $d=1.5$ is located at $x=y=z=8.5$.

The plane vortex configuration attracts a zero mode just like the spherical
vortex, according to their topological charge with modulus $|Q|=1$.
\begin{figure}[h!]
\centering
a)\includegraphics[width=.3\columnwidth]{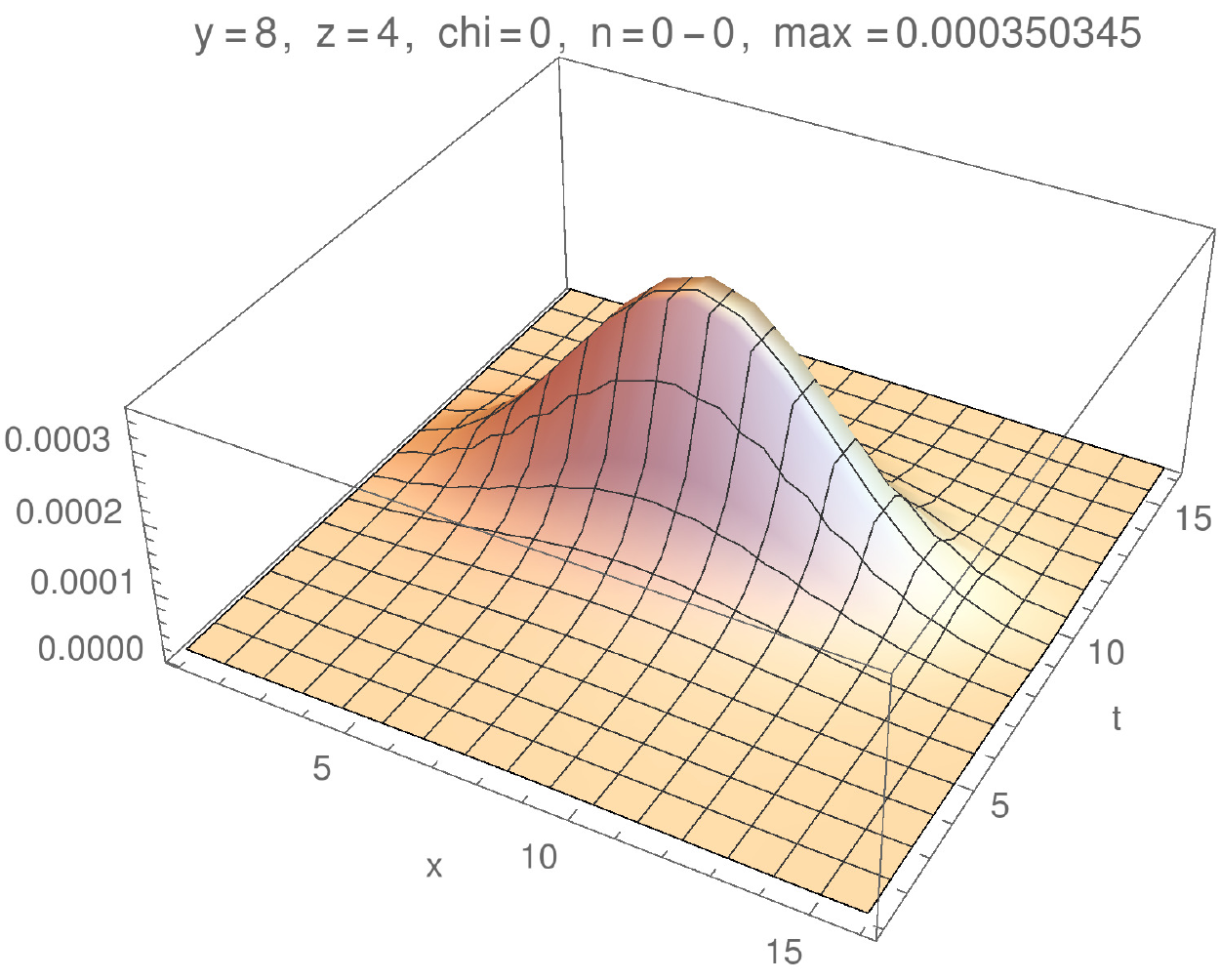}
b)\includegraphics[width=.3\columnwidth]{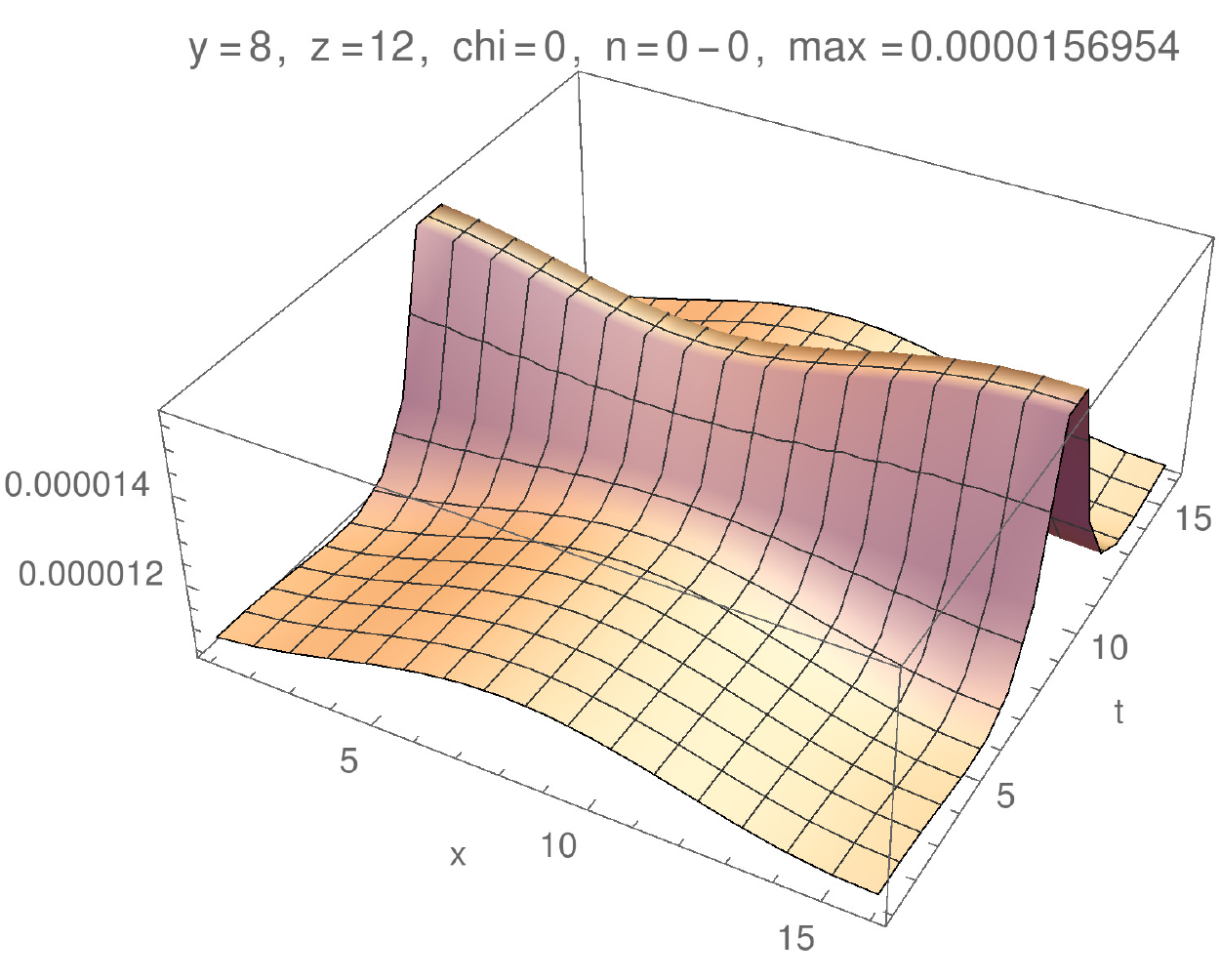}
c)\includegraphics[width=.3\columnwidth]{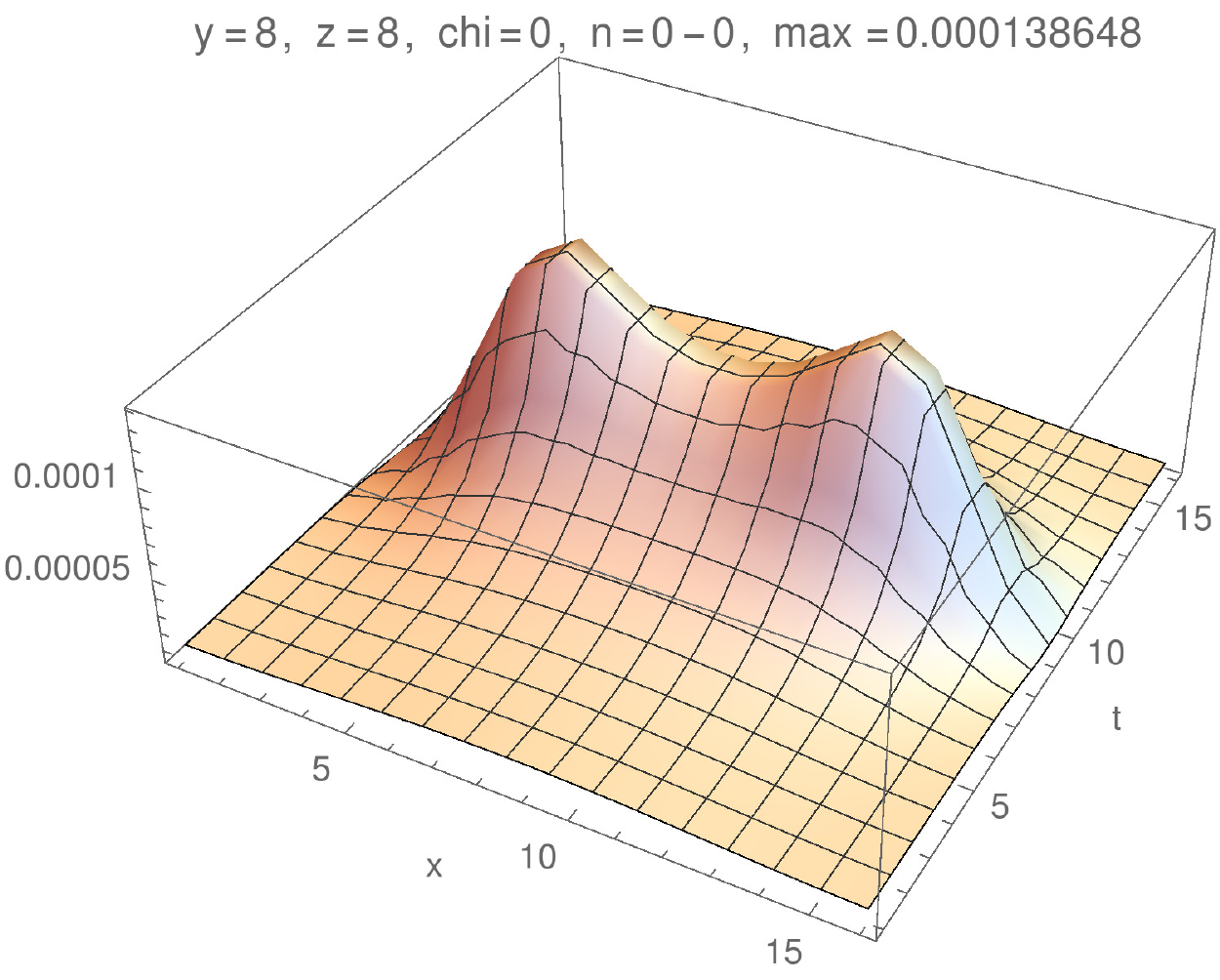}
\caption{Chiral densities of overlap eigenmodes; a) zero mode $\rho$\#0(left) for the colorful region of anti-parallel colorful plane vortices; b) the same as a) but for unicolor region; c) the same as a) but for a spherical vortex with $Q=-1$.}
\label{chiral1}
\end{figure}
In Fig.~\ref{chiral1} we show that the chiral densities of the zero modes for
anti-parallel colorful plane vortices and spherical vortices with $Q=-1$. As
these densities are axially symmetric in the $xy$-plane around a center we plot
$xt$-slices. We select the colorful and the unicolor regions by fixing the
$z$-coordinate appropriately. The plot titles of the density plots give the $y$-
and $z$-coordinates of the $xt$-slices, the chiral density ({\it i.e.}, "chi=0"
means we plot $\rho_5$), the number of plotted modes ("n=0-0" means we plot
$\rho\#0$) and the maximal density in the plotted area ("max=..."). The boundary
conditions for fermion fields and the parameters for the colorful plane vortices
and the spherical vortices are the same as in Fig.~\ref{overlap}.

The response of fermions to the plane vortices is squeezed in time direction, since the vortices are localized in a single time slice (fast transition). For the plane and spherical colorful vortices the chiral density of the zero mode peaks at the center of the colorful region ($x=y=8$), see Fig.~\ref{overlap}. The value of chiral density is positive {\it i.e.} the chirality of the zero mode is right handed as expected from the index theorem. For the spherical vortex we observe two peaks because we have two colorful regions along the $x$-direction at $y=z=8$. In the unicolor region of plane vortices we don't have any topological charge density, nevertheless we observe some influence of the topological charge density of the colorful region on the chiral density of the unicolor region.
\begin{figure}[h!]
\centering
a)\includegraphics[width=0.46\columnwidth]{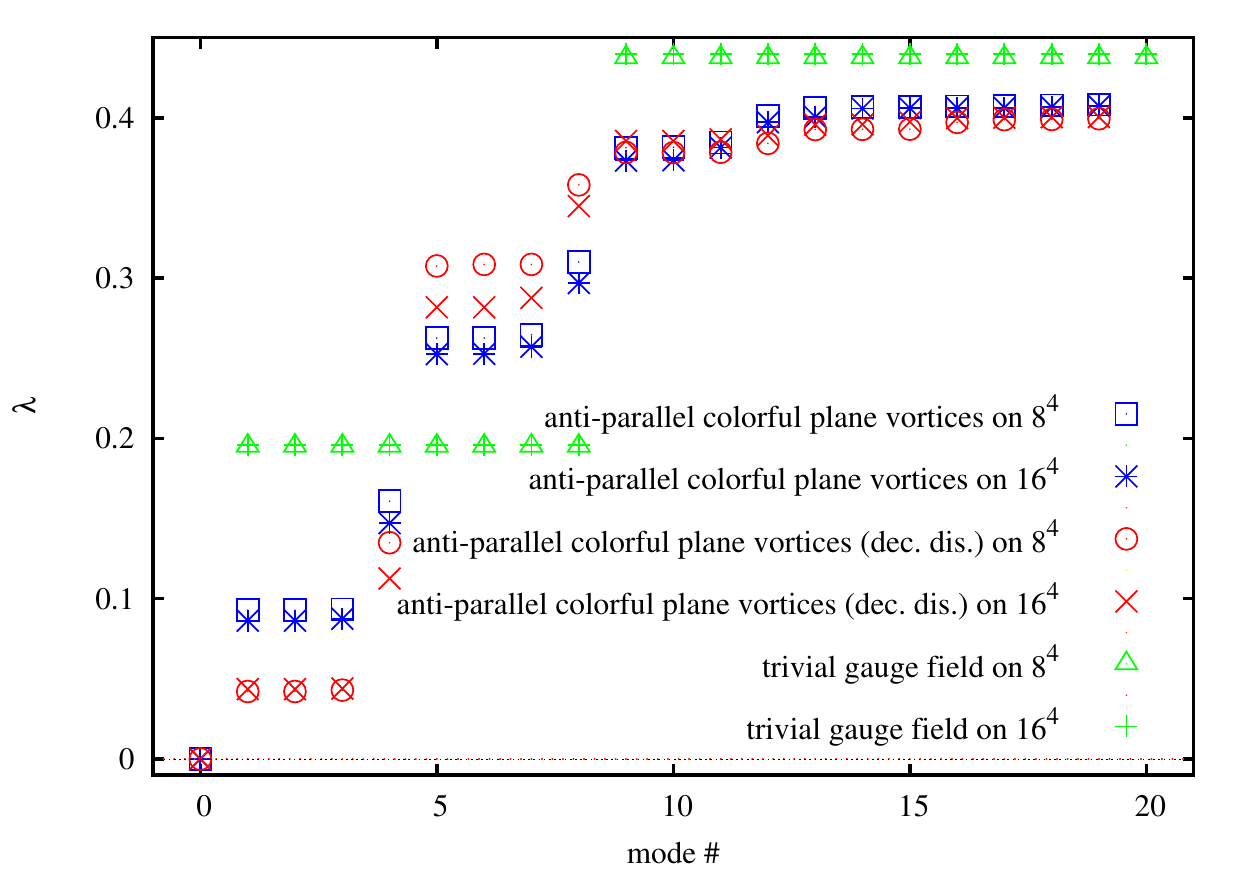}
b)\includegraphics[width=0.46\columnwidth]{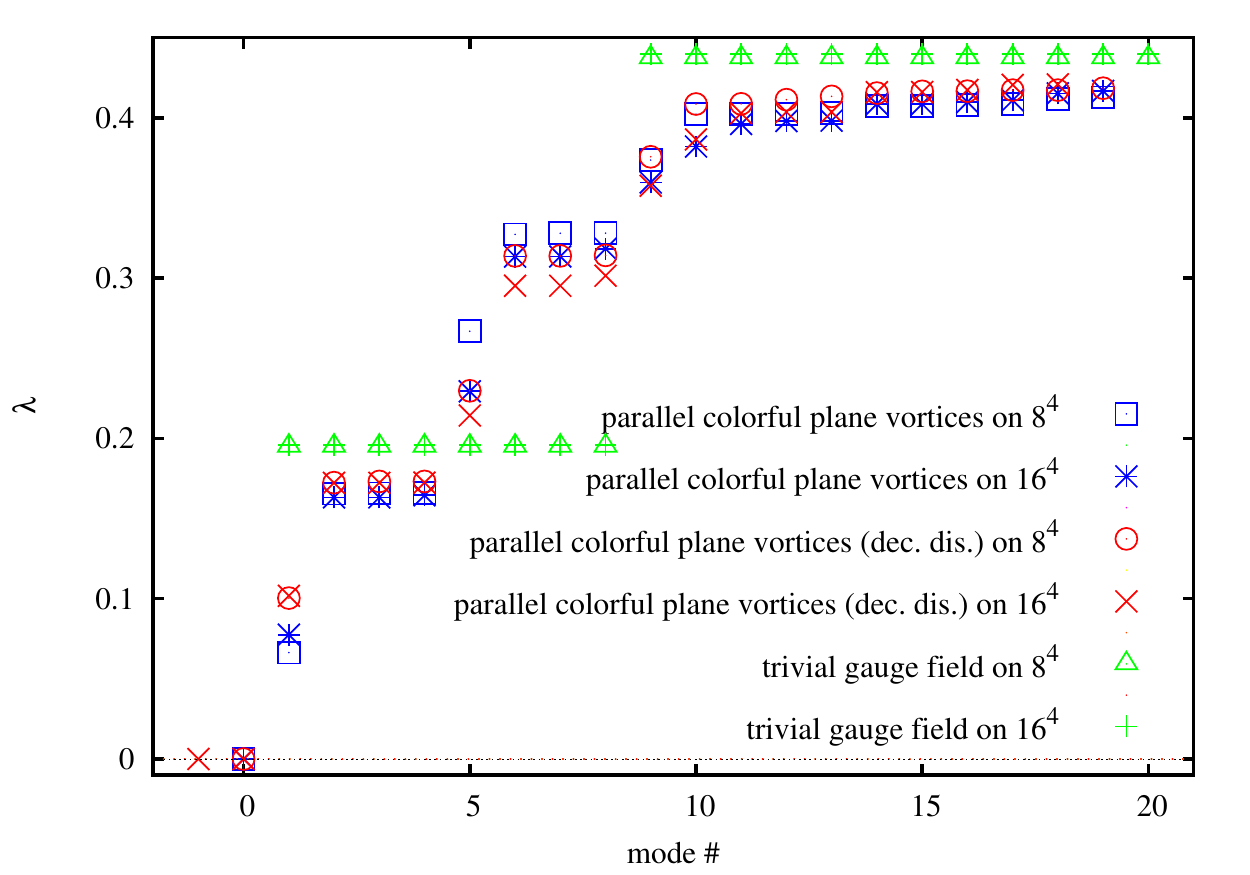}
\caption{The lowest overlap eigenvalues on $8^4$ and $16^4$ lattices a) for anti-parallel plane vortices  b) for parallel plane vortices compared to the eigenvalues of the free (overlap) Dirac operator, which are scaling inversely proportional to the linear extent of the lattice. To simplify the comparison of eigenvalues for the two different lattice sizes the smaller eigenvalues of the $16^4$ lattice are multiplied by $2$.}
\label{eigen2}
\end{figure}

As shown in Fig.~\ref{overlap}, for colorful plane vortices, we also get four
low lying eigenmodes with smaller eigenvalues than the ones of the lowest
eigenvectors for the trivial gauge field. It is interesting to mention that these
low lying modes can not be removed by changing the boundary conditions while
these low lying modes can be removed by appropriate boundary condition for
unicolor plane vortices. In Fig.~\ref{eigen2}, we show the behavior of these low
lying modes for parallel and anti-parallel colorful plane vortices by changing
the lattice size and the distance between two sheets of the colorful plane
vortices. The boundary conditions for fermion fields and the parameters for
colorful plane vortices are again the same as in Fig.~\ref{overlap}. The
decreased distance between the two sheets is half of the initial distance.
However, to avoid the overlap of two plane vortices we also decrease the
thickness to half the initial thickness.

For anti-parallel plane vortices, after increasing the lattice size the
eigenvalues of the four low lying eigenmodes decrease. Therefore we do not
expect them to approach the eigenvalues of the trivial gauge field in the
continuum limit and identify them as near zero modes. Interestingly, with decreasing distance between the vortex sheets these eigenvalues decrease even further.
For parallel plane vortices, the four low-lying eigenvalues approach trivial
ones for increasing lattice size and decreasing distance between the vortex
sheets, hence we cannot identify near zero modes.

We conclude that anti-parallel vortex pairs with color structure may contribute to a finite density of near-zero modes leading to chiral symmetry via the Banks-Casher relation.


\section{Conclusion}\label{sec:conclusion}
We have investigated plane vortex configurations where one of the vortex sheets
has a topological non-trivial color structure. To define this colorful vortex
configuration we start from a configuration with non-trivial temporal links
$U_4$ and trivial spatial links $U_i$. For such a configuration, one gets
vanishing gluonic topological charge but the difference of the left and right
chiral zeromodes is $n_--n_+=-1$ and the index theorem with $Q=n_--n_+$ is not
fulfilled.  The discrepancy is simply a discretization effect. In continuum,
this configuration is a fast transition of a vacuum with winding number
$N_{w_1}=0$ to a vacuum with winding number $N_{w_2}=-1$. Therefore, in the
continuum limit this vortex configuration has a topological charge
$Q=N_{w2}-N_{w1}=-1$. After smoothing this continuum object in temporal
direction and putting it onto the lattice, the topological charge converges to
$-1$. As expected, cooling or smearing of this generalized vortex leads to an
anti-instanton configuration with four dimensional spherical symmetry and its
center located in the center of the rotational symmetric color structure of the
colorful plane vortex. In Monte Carlo configurations we do not,
of course, find perfectly flat or spherical vortices, as one does not find
perfect instantons, which are only recovered after cooling or smearing. 
The general picture of topological charge from vortex 
intersections, writhing points and various color structure contributions, e.g.
spherical or the colorful plane vortex configuration presented here, or
instantons can provide a general picture of $\chi$SB: any source of
topological charge can attract (would-be) zero modes and produce a finite
density of near-zero modes leading to chiral symmetry breaking via the
Banks-Casher relation. In fact, using the overlap Dirac operator, we have calculated eigenmodes in the background of this colorful plane vortex configuration. Due to the index
theorem, this configuration attracts one zero mode which is concentrated at the
colorful vortex sheet. In addition to this zero mode we find for anti-parallel
plane vortices four low lying modes which persist regardless of the boundary
conditions, while they can be removed by antiperiodic boundary condition for
unicolor plane vortices. With increasing lattice size the eigenvalues of these
four low lying modes are decreasing substantially and we identify them as near
zero modes. We have given an additional demonstration that in addition to intersections and
writhing points the color structure of vortices can contribute to the
topological charge if the vortices are thick, smoothed over several lattice
slices, and that such vortices contribute to the density of near zero modes,
which may lead to chiral symmetry breaking via the Banks-Casher relation.

\acknowledgments{Seyed Mohsen Hosseini Nejad would like to thank his supervisor Sedigheh Deldar for her great support. We are grateful to {\v Stefan} {Olejn\'\i k} for interesting discussions. Roman H\"ollwieser was supported by the Erwin Schr\"odinger Fellowship program of the Austrian Science Fund FWF (``Fonds zur F\"orderung der wissenschaftlichen Forschung'') under Contract No. J3425-N27.}

\bibliographystyle{utphys}
\bibliography{../literatur}

\end{document}